\documentclass[aps,prb,twocolumn]{revtex4}

\def\bea{\begin{eqnarray}}
\def\eea{\end{eqnarray}}
\def\beq{\begin{equation}}
\def\eeq{\end{equation}}

\usepackage{epsfig,color}
\usepackage{dcolumn}
\usepackage{bm}
\usepackage{graphicx}
\usepackage{epstopdf} 
\begin{document}
\bibliographystyle{apsrev}
\title{Magnetic resonance in the cuprates -- exciton, plasmon, or $\pi-$mode }
\author{Zhihao Hao}
\affiliation{Department of Physics and Astronomy, Johns Hopkins University,
3400 N. Charles St., Baltimore, Maryland 21218, USA}
\author{A. V. Chubukov}
\affiliation{Department of Physics, University of Wisconsin,
1150 University Ave., Madison, Wisconsin 53706, USA}
\begin{abstract}
We re-analyzed the issue whether the resonance peak observed in neutron scattering experiments on the cuprates is an exciton, a $\pi-$resonance, or a magnetic plasmon. We considered a toy model with on-cite Hubbard $U$
 and nearest-neighbor interactions in both charge and spin channels. We found that the resonance is predominantly an exciton, even if  magnetic interaction is absent and $d-$wave pairing originates from attractive density-density interaction. Our results indicate that one cannot distinguish between spin and charge-mediated pairing interactions by just looking at the resonance peak in the dynamic spin susceptibility.
\end{abstract}
\maketitle
{\it Introduction.}~~The origin of the $(\pi,\pi)$ spin resonance in the cuprates continue to attract interest of the high-$T_c$ community. The resonance has been observed in four
 different classes of high-$T_c$ compounds - YBCO, $Bi2212$, $Tl2201$ and $Hg1201$~\cite{neutrons},  and the doping variation of its energy follows closely the doping dependence of $T_c$. Magnetic resonances have been also recently observed in heavy-fermion materials~\cite{hf} and in $Fe$-pnictides~\cite{fe}.

It is widely accepted that the
 resonance at ${\bf Q} = (\pi,\pi)$ is a feedback from the opening of a $d-$wave pairing
gap in the fermionic spectrum
 There is no consensus, however, about the driving force.  The simplest and most transparent idea put forward by various groups~\cite{feedback}
 is that the neutron
resonance is a spin exciton, that is, a resonance mode in the spin
response function, which emerges
due to an attractive  residual spin interaction between quasiparticles
 in a $d-$wave superconductor.  To obtain this mode, one can either compute the susceptibility within the low-energy model with spin interaction only
(no charge component)~\cite{acs}, or just  calculate  the spin susceptibility  within a conventional RPA for the underlying Hubbard model~\cite{rpa} --
 either way one
 obtains a   $\delta-$functional excitonic peak at a finite frequency below $2\Delta$, where
 $\Delta$ is the amplitude of gap at  ``hot spots'' -- $k_F$ points separated by ${\bf Q}$.

This simple approach, however, is incomplete as it
 neglects the fact that in a $d-$wave superconductor
 the staggered particle-hole, charge 0
  spin variable ${\bf S}^a_Q = (1/2\sqrt{N}) \sum_k c^\dagger_{k,\alpha} \sigma^a_{\alpha \beta} c_{k+Q,\beta}$ is mixed with the
staggered $d-$wave particle-particle charge $\pm 2$ variables
 $\pi^a_Q$ and $(\pi^a_Q)^*$, where $\pi^a_Q = (1/\sqrt{N})
\sum_k d_k c_{k,\alpha} (\sigma^a \sigma^y)_{\alpha \beta} c_{k+Q,\beta}$, and $d_k = cos k_x - cos k_y$ (Refs. [\onlinecite{zhang,oleg,lee}]).
Diagrammatically, mixed $<S \pi>$ response function is given by
$d_k G_k F_{k+Q}$ bubbles made out of normal ($G$)  and anomalous ($F$) Green's functions~\cite{oleg}. Such terms are finite in a $d-$wave superconductor at $\omega \neq 0$.

  Because spin and $\pi$ responses are coupled, the full spin response function is obtained by
solving the full $3\times 3$  set of coupled
 generalized RPA equations for $<SS>, <\pi,\pi>$ and $<S\pi>$ correlators (Ref.~\cite{comm_1}).
 As a consequence, the resonance mode emerges simultaneously in spin
 and $\pi$ channels, and  for the case when both $s$ and $\pi$ resonances are present,
 its location $\omega = \omega_{res}$
is in general the solution of
\beq
\chi^{-1}_s (\omega) \chi^{-1}_\pi (\omega) - \omega^2 C^2_\omega =0
\label{1}
\eeq
where $\chi^{-1}_s \propto (\omega - \omega_{exc})$ and $\chi^{-1}_\pi \propto (\omega - \omega_\pi)$
 are inverse RPA susceptibilities in $s$ and $\pi$ channels, each resonating at its own frequency, and
$\omega~ C_\omega$ is the mixing $GF$ term~\cite{comm}.
 If $C_\omega=0$,
 $s$ and $\pi$ channels are decoupled, and the spin and $\pi$
resonances occur  at $\omega_{exc}$ and $\omega_\pi$, respectively,
 and is are not affected by each other. In general, however, the resonance frequency $\omega_{res}$ is the solution of (\ref{1}), and the full
spin and $\pi$ susceptibilities near the resonance are given by
 $\chi_s = Z_s/(\omega - \omega_{res}), ~\chi_{\pi} = Z_{\pi} /(\omega - \omega_{res})$ [we normalize $Z$ such that for $C_\omega =0$, $Z_s =1, Z_\pi =0$ at $\omega = \omega_{exc}$, and $Z_\pi =1, Z_s =0$ at $\omega = \omega_\pi$].

Eq. (\ref{1}) shows that, in general, there are three possibilities
 for the neutron resonance. It can be an exciton, which is the case when
$\omega_{res} \approx \omega_{exc}$ and $Z_s >> Z_{\pi}$ (Refs. [\onlinecite{feedback,acs,rpa}]).
It can also be $\pi$ resonance~\cite{zhang},  which holds when $\omega_{res} \approx \omega_\pi$, and $Z_\pi >> Z_s$.  And, finally, it can be a magnetic plasmon~\cite{lee}, which is the case when
$\chi_s (\omega)$ and $\chi_{\pi} (\omega)$ weakly depend on frequency, and the resonance emerges due to the mixing between the two channels. In this last case, $\omega_{res} \approx
 (\chi^{-1}_s (0) \chi^{-1}_\pi (0)/C_0^2)^{1/2}$ and is generally different from both $\omega_{exc}$ and $\omega_\pi$.

Another issue which we consider is the relation between the
 spin resonance and the ``glue'' for a $d-$wave superconductivity, at least at and above optimal doping, where
 the system falls into moderate coupling regime~\cite{timusk}.
 The pairing can be magnetically-mediated~\cite{acs}, or it can  be mediated by a $d-$wave attraction in the charge channel~\cite{italy}. We will analyze
 whether the location and the residue of the resonance can distinguish between the two cases.

We follow earlier works, use BCS approximation, and model the attractive spin-dependent interaction by nearest-neighbor Heisenberg exchange term $J>0$, (Ref. \cite{oleg}), and a $d-$wave interaction in the charge channel  by nearest-neighbor density-density interaction $V$ (Ref. \cite{lee}).
For a repulsive charge interaction ($V >0$),
there is no $\pi$ resonance (i.e., no pole in $\chi_\pi$ below $2\Delta$)
 ~\cite{oleg}, hence the neutron resonance can be either an exciton or a plasmon. For negative $V$, both $\chi_s$  and $\chi_\pi$ have poles below $2\Delta$,
 and the resonance can be an exciton, a $\pi$-resonance, or a plasmon.
To distinguish between them, we will solve the
 full $3 \times 3$ matrix
 equation for $\chi$, compare the residues in spin and $\pi$ channels,
(this should show whether the resonance is an exciton or a $\pi$ resonance),
 and  also compare the location of the pole with $\omega_{exc}$ (this should show whether or not the resonance is a plasmon). We follow earlier work~\cite{lee}
 and  require that the value of a $d-$wave gap should agree with
 ARPES experiments~\cite{arpes}

Our results show that, to a surprisingly good accuracy, the
 resonance remains an exciton no matter whether the pairing is in the spin or in the charge channel. For both cases, we found that  neither $\pi-$resonance
 nor the mixture between $s$ and $\pi$ channels affect the location and the residue of the resonance residue in any substantial way, although the corrections
 due to mixture are larger for the case $|V| >> J$.
Furthermore, we find that $Z_s$ can be large enough and
the resonance frequency can be the experimental $40 meV$  without placing the system too close to an antiferromagnetic instability.

Our results disagree with the idea about the dominance of the $\pi-$ resonance~\cite{zhang}, and also somewhat disagree with the recent study~\cite{lee}
which associated the resonance with a plasmon rather than an exciton.
 That work also found that the study of the resonance can distinguish between
spin and charge mechanisms in favor of the former. We, on the contrary,
 found that the resonance is only weakly sensitive to the form of the pairing glue. Still, we find, in agreement with Ref. [\onlinecite{lee}], that
 the mixing between spin and $\pi$ channels is not completely negligible and has to be taking into account in quantitative studies of the cuprates. 
 We caution that, for the charge-mediated pairing, the results stronly depend on the magnitude of nearest-neighbor attraction $V$. We have chosen $|V|$ which yields a BCS gap of $35 meV$. This $|V|$ turns out to be too small to give rise to $\pi$ resonance.  For larger $|V|$, the structure of $\chi_s (\omega)$ will  differ more from an exciton. 

{\it The model.}~~~~
We consider the same model as Lee et al (Ref. ~\onlinecite{lee}),
 with on-site Hubbard repulsion and nearest-neighbor density-density and spin-spin interactions,
\begin{equation}\label{themicroscopicmodel}
H = \sum_{\mathbf{k},\sigma}(\epsilon_\mathbf{k}-\mu)
a^{\dagger}_{\mathbf{k}\sigma}a_{\mathbf{k}\sigma} +  \sum_i U \, n_{i\uparrow} n_{i\downarrow} + \sum_{\langle ij \rangle}(V \, n_i n_j
             + J \, \mathbf{S}_i\cdot\mathbf{S}_j),
\end{equation}
where $ \epsilon_{\mathbf{k}}=-2t(\cos k_x+\cos k_y )
    -4t^{\prime}\cos k_x \cos k_y$,  $n_{i\sigma} = c^\dagger_{i\sigma} c_{i\sigma}$ and ${\mathbf S}_i =  (1/2) c^\dagger_{i\sigma}{\bf \sigma}_i c_{i\sigma}$ are the particle and spin operators on site $i$ (each interaction is counted once). A similar model but without $U$ term has been earlier considered by Norman and one of us~\cite{oleg}.

The soft modes of the system are singlet pairs on nearest-neighbor bonds,
$\psi_{ij} = a_{i\alpha} \sigma^y_{\alpha\beta}a_{j\beta}$, spin fluctuations
${\vec S}_{ij} = (1/2) a^\dagger_{i\alpha} {\vec \sigma}_{\alpha\beta}a_{j\beta}$,
and triplet pairs ${\vec \pi}_{ij} =  a_{i\alpha}
({\vec \sigma} \sigma^y)_{\alpha \beta}  a_{j\beta}$.
The gap  $\Delta_\mathbf{k} = \Delta g_\mathbf{k}$
with $g_\mathbf{k} = (\cos{k_x}-\cos{k_y})/2$ is determined from the standard equation
\begin{equation}\label{BCSgap}
-\frac{V_\psi}{2}\int\frac{d^2 k}{(2\pi)^2}
\frac{g_{\mathbf{k}}^2}{\sqrt{(\epsilon_{\mathbf{k}}-\mu)^2+\Delta^2
g_{\mathbf{k}}^2}} = 1.
\end{equation}
where $V_\psi = V - 3 J/4$.
Choosing  $x=0.12$ ($\mu = -0.94 t$), $t'/t=-0.3$ and  $t=0.433$ eV to
 match the observed shape of the Fermi surface and the nodal Fermi velocity~
\cite{nodal}, and setting the maximum gap to be $\Delta =35 meV$,
we find $V_\psi = V - 3J/4 = -0.60t$, in agreement with ~\cite{lee}.

\begin{figure}
\includegraphics[width=0.9\columnwidth]{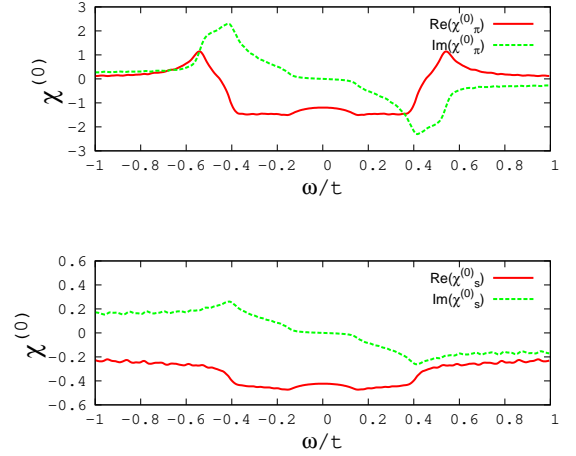}
\caption{The bare susceptibilities in the spin and $\pi$ channels,
$\chi^0_{s} (\omega)$ and  
$\chi^0_{\pi} (\omega) = \chi^0_{11}$, respectively 
(in units of $1/t$)
The susceptibility for the amplitude $\pi$ channel
 $\chi^0_{22}$ is smaller and featureless. 
 Note the difference in vertical scales -- $\chi^0_\pi$ is larger than $\chi^0_s$. 
 }\label{bare}
\end{figure}

{\it Generalized RPA susceptibilities}~~~
We  use $V_\psi$ and $\Delta$ as inputs and compute
 dynamic susceptibilities within a generalized RPA scheme
which, we remind, takes into accunt the fact that a particle-hole and a particle-particle channels are mixed in the presence
of a charged condensate of Cooper pairs.  The derivation of the generalized RPA equations is rather straightforward and has been described
before~\cite{zhang,oleg,lee}.
Because of $SU(2)$ spin symmetry, it is sufficient to probe only
 one spin component, e.g., restrict, in momentum space with
$A_0 (q) = S^+ (q)$, $A_1 (q) = (\pi^y (q) +i \pi^x (q))/2$, and $A_2 (q)= A^*_1 (q)$.
These three operators create bosonic excitations with the same
 momentum and spin $S_z =1$, but with different charges, $0$ and $\pm 2$, respectively. For definiteness, we restrict with antiferromagnetic $q=Q = (\pi,\pi)$.

Generalized RPA equations relate bare and full
 susceptibilities:
\begin{equation}
\chi_{a,b}(\omega) = \chi^0_{a,b}(\omega)
+ \chi^0_{a,c}(\omega) \Gamma_{c,d} \chi_{d,b}(\omega),
\end{equation}
where $a,b =0,1,2$, and
$\chi^0_{a,b}(\omega)$ are linear response functions for the
noninteracting system  -- the Fourier transforms of
$ -i \theta (t) \langle [A_\alpha(t),A^\dagger_\beta(0)] \rangle$, where the averaging is over free fermion ground state. Diagrammatically, $\chi^0_{aa}
(\omega)$ are the convolutions of $GG$ and $FF$ terms, while 
 non-diagonal $\chi^0_{a,b} (\omega)$ are $GF$ terms.
 All 9 elements of $\chi^0_{a,b}(\omega)$ are nonzero, but non-diagonal terms 
 $\chi^{0}_{01}$ and $\chi^0_{12}$  vanish at zero frequency, and $\chi^0_{02}$ is nonzero only because of the curvature of the fermionic dispersion near $k_F$. It is convenient to
 rotate the basis in $\pi, \pi^*$ plane and introduce, instead of $A_{1,2}$,  ${\bar A}_{1,2} (A_1 \mp A_2)/\sqrt{2}$ (same notations have been used in earlier works ~\cite{oleg,lee}). In this basis, the interaction
matrix $\Gamma_{a,b}$ is diagonal due to charge conservation, and its diagonal elements are $V_0 = -U -2J, V_{1,2} = (V+ J/4)/2$, 
 i.e., without mixing $\chi_s = \chi^0_s/(1 - V_0 \chi^0_s),~\chi_\pi = \chi^0_\pi/(1 - V_{1,2} \chi^0_\pi)$.
 
In Fig. \ref{bare} we show real and imaginary parts of the  
 bare $\chi^{0}_s (\omega) = \chi^0_{00} (\omega)$ and $\chi^{0}_{\pi} (\omega) =\chi^0_{11} (\omega)$ (a phase mode of the $\pi$ field). The bare $\chi^0_{22}$ is smaller than $\chi^0_{11}$.
 Note that in our notations, static$\chi^{0}_{s,\pi}$ are negative.
 Sharp features  at $\pm 0.16t$
( $\pm 70 meV$)  are $2\Delta$ effects,
the features at higher energies  are Van Hove singularities.

{\it The Results.}~~~
We present the results for the  two extreme cases
 cases  $V=0$ and $J=0$. In the first case, $d-$wave superconductivity is
magnetically-mediated and  $V_\psi = -3J/4$. In the second it
 emerges due to an attraction in the charge channel, and $V_\psi = V$.
For both cases, we used on-site Hubbard $U$ as an extra parameter that drives the system towards an antiferromagnetic instability and brings the resonance frequency down.

\begin{figure}
\includegraphics[width=0.9\columnwidth]{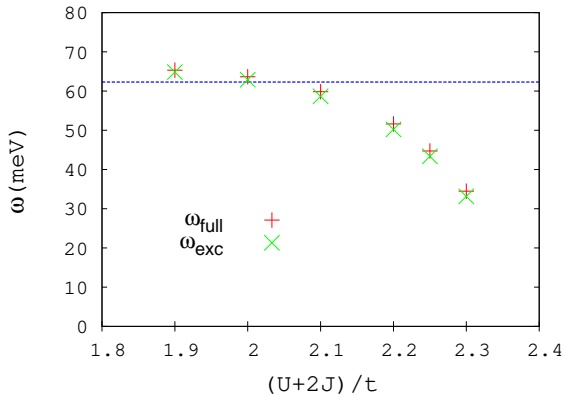}
\caption{The resonance positions for $V=0$ and $J =0.8t$ for
different $U$. $\omega_{full}$ is the solution of the full $3\times3$ set, $\omega_{exc}$ is the energy of a spin exciton.  The blue dashed line is the edge of
  two particle continuum.  
We verified analytically that for $V_{12} >0$ (this case), $\omega_{full} > \omega_{exc}$.}
\label{resonance199}
\end{figure}
\begin{figure}
\includegraphics[width=0.9\columnwidth]{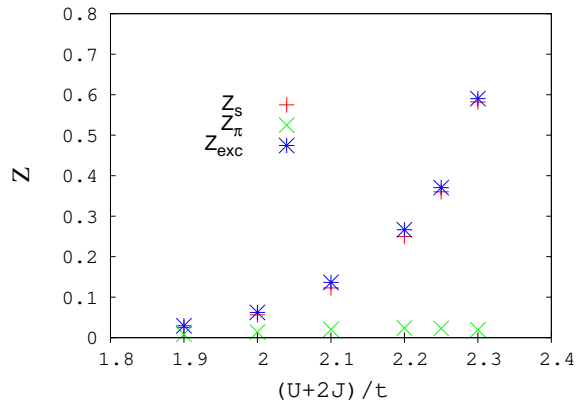}\\
\caption{The residues $Z_s$, $Z_\pi$ from the full $3 \times 3$ set,
 and the residue of an exciton $Z_{exc}$ for different $U$.}\label{residue199}
\end{figure}

For the first case, $V_0 = -(U + 1.6 t), V_{1,2} = 0.1t >0$, and
 $\pi-$susceptibility taken alone only
 develops an anti-resonance above the upper edge of two-hole
continuum~\cite{oleg}. The issue for this case is
 whether the resonance is an exciton or a plasmon.
In Fig. ~\ref{resonance199} we present the results of our calculations of the resonance frequency using the full $3\times 3$ set and compare them with the
  RPA result for spin-only channel.
 We see that the energies match nearly perfectly, which we believe is
 a strong indication that the resonance is indeed the exciton. 
On a more careful look, we found that a near-perfect match is not the consequence of the small mixing amplitude ($C_\omega$ in (\ref{1}) is roughly $1.3 t$) but rather the consequence of the fact that $\omega =50 meV$ is only $0.12 t$.
In Fig. ~\ref{residue199} we plot the residues of the resonance
in spin and $\pi$ channels, $Z_s$ and $Z_\pi$, respectively, together with the residue of a pure exciton, $Z_{exc}$, which we  obtained by
 eliminating the mixing between  spin and $\pi$ channels.
 We used a finite broadening $\gamma =0.002$ which
explains why $Z<1$  even for the case of a pure exciton.   We see that
for all $U$, the residue of the resonance is much
 larger in the spin channel than in the $\pi$ channel. If the resonance was a plasmon, the residue in the spin and $\pi$ channels would be comparable.
As an independent check, we solved
a  $3\times 3$ set with diagonal $\chi^0_{aa} (\omega)$ replaced by their static values. The solution in this case would be a plasmon (see Eq. (\ref{1})),
 but we didn't find a resonance.

Fig. ~\ref{resonance199}  also shows that the resonance shifts down
from $2\Delta$ (and becomes strong)
when $|V_0|$ exceeds roughly 80\% of the critical $|V_0| = 2.38 t$.
 beyond which antiferromagnetic order emerges.
 Using $\xi \propto (\xi/a)^2$, where $\xi$ is the correlation length
 and $a$ is interatomic spacing, we find that this corresponds to
 $\xi \sim 2.5a$, while $\xi \sim 3.7 a$ is necessary for the resonance frequency $\omega_{res}$ to be $40 meV$.

\begin{figure}
\includegraphics[width=0.9\columnwidth]{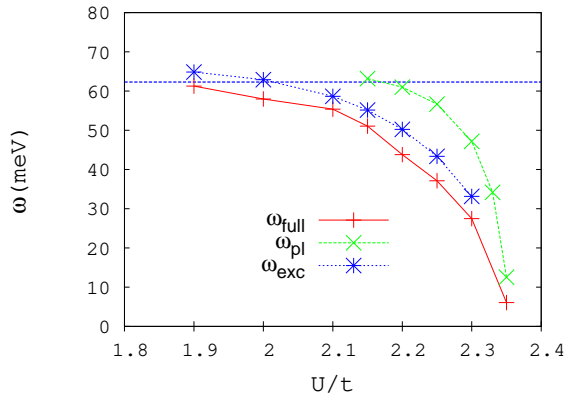}
\caption{The energy of the resonance from the solution of the full $3
\times 3$ set ($\omega_{full}$), and the energies of an exciton and a lasmon 
($\omega_{exc}$ and $\omega_{pl}$) for the case of nearest-neighbor charge interaction $J=0, V =-0.6 t$ as functions of $U$ (see text). 
 The blue dashed line is the edge
of the two particle continuum. 
In this case, $V_{12} <0$, and $\omega_{full} < \omega_{res}$.
}\label{resonance597}
\end{figure}

\begin{figure}
  \includegraphics[width=0.9\columnwidth]{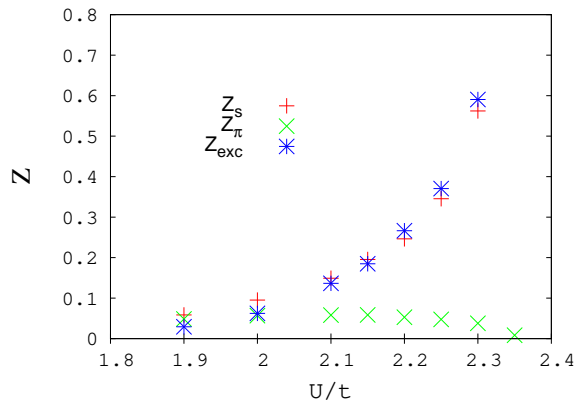}\\
  \caption{The residues $Z_s$, $Z_\pi$, and
 $Z_{exc}$ for the case $J=0, V =-0.6 t$ for different $U$.}\label{residue597}
\end{figure}

The results for the second case, $J=0, V =-0.6 t$, i.e., $V_0 = -U$,
$V_{1,2} = -0.3 t$,   are presented in Figs.
\ref{resonance597} and ~\ref{residue597}. Now
  $\pi$-channel becomes are attractive,
 and exciton, plasmon, and $\pi$ resonance are all competing for  the dominant contribution to the resonance in the full spin susceptibility.
 The three sets of points in
Fig. \ref{resonance597} are the solution of the full $3\times 3$ set and
 two approximate sets in which we (i) considered the spin channel only (the resonance is an exciton)  and (ii)
 approximated diagonal $\chi^{0}_{aa} (\omega)$ by their static values
 (the resonance is a plasmon).
We see that the position of the actual resonance (the full solution) is rather close to the position of the spin exciton and the two follow the same trend with $U$, although there is a clearly visible difference of about $5-10\%$.
 The plasmon has different dependence on $U$, and is located below $2\Delta$ only in a narrow range of $U$. 
  Note that $\omega_{full} < \omega_{exc}, \omega_{pl}$, as it indeed should be (see Eq. (\ref{1})). In Fig. ~\ref{residue597} we show the residues of the spin and
 $\pi$ components of the full susceptibilities near $\omega_{res}$ and
 compare them with the residue of a pure spin exciton (the case (i) above).
 We see  that, when the resonance shifts below $2\Delta$ and becomes measurable, its residue in the spin channel  is larger than in the $\pi$ channel and
 practically coincides with the residue of an exciton.
Note that the same $\xi/a \sim 3.7$ as in the first case is required for
 the resonance to be at $40 meV$.


These results  imply that, even if the pairing is due to an
attractive nearest neighbor density-density interaction,
the resonance in the spin susceptibility
 still has  predominantly excitonic character.
We caution, however,
the absence of a substantial $\pi$ component of the neutron resonance
is as a numerical rather than a fundamental effect. Namely,
 for $V = -0.6 eV$ extracted from the  gap equation, $\pi-$resonance  
 in the absence of broadeing is close to $2\Delta$, and  a small broadening washes it out. We also note that a plasmon  does exist
 in this case, in agreement with Ref. ~\cite{lee}, but in a
narrow range of $U$ near an antiferromagnetic instability.

We also performed calculations for several  $J$ and $V$ in between the two limits, still keeping $V_\psi = V-3J/4 = -0.6 t$ to match the gap value.
For all cases we  found that the resonance is predominantly an
 exciton.  
The situation canges if we abandon BCS gap equation and take larger $|V|$. 
The larger $|V|$ is, the stroner the resonance differs the exciton.

To conclude, in this paper we re-analyzed  whether the resonance peak observed in neutron scattering experiments on the cuprates is an exciton, a $\pi-$resonance, or a magnetic plasmon. We considered a model with on-cite Hubbard $U$
 and nearest-neighbor interaction in both charge and spin channels and
  found that the resonance is predominantly an exciton,
even if  $d-$wave pairing originates from attractive density-density interaction rather than spin-spin interaction.
 Our results indicate that one cannot distinguish between 
  spin and charge-
-mediated pairings  by looking at the resonance peak in the spin susceptibility.
Other probes like e.g., dispersion anomalies~\cite{norm} or
Raman scattering~\cite{raman} are more useful in this regard.

We acknowledge with thanks useful discussions with Wei-Cheng Lee, A.H. MacDonald, O. Tchernyshov, and M.R. Norman.  This work  was supported by NSF-DMR-0520491 (Z. H), 
and by  NSF-DMR 0604406 (A.V. Ch). 
  We thank Wei-Cheng Lee and A.H. MacDonald for careful reading of the manuscript and useful remarks.


\begin{thebibliography}{99}
\bibitem{comm_1} In general, one has to include also $A_{1g}$ spin component $ {\tilde S}^a_Q (1/2\sqrt{N}) \sum_k s^+_k c^\dagger_{k,\alpha} \sigma^a_{\alpha \beta} c_{k+Q,\beta}$, where $s^+_k = \cos k_x + cos k_y$,
but $<{\tilde S} {\tilde S}>$ correlator is very small numerically~\protect\cite{lee}.
\bibitem{comm} The actual equation for the resonance frequency (the solution
 of $3\times3$) set is a bit more involved because $\pi^a_Q$ is a complex field.
\end{thebibliography}
\end{document}